\documentclass{synmet}
\usepackage{graphicx}
\usepackage{amssymb}
\usepackage{epsfig}

\begin{document}

\begin{frontmatter}


\title{Electron-hole versus exciton delocalization in conjugated polymers: the
role of topology}
\author[label1]{S. Dallakyan,}
\author[label2]{M. Chandross,}
\author[label1]{and S. Mazumdar}
\address[label1]{Department of Physics, University of Arizona, Tucson, AZ, 
	85721, USA}
\address[label2]{Sandia National Laboratories, Albuquerque, NM 87185, USA}

\begin{abstract}
There is currently a great need for solid state lasers that emit in the infrared.
Whether or not conjugated polymers that emit in the IR can be synthesized is an
interesting theoretical challenge. We show that the requirement for such a
material is that the exciton delocalization in the system be large, such that
the optical gap is small. We develop a theory of exciton delocalization in
conjugated polymers, and show that the extent of this can be predicted from
the topology of the conjugated polymer in question. We determine the precise
structural characteristics that would be necessary for light emission in the
IR.
\end{abstract}

\begin{keyword}
Polyacetylene and derivatives \sep Semi-empirical models and model 
calculations \sep Light sources.

\end{keyword}

\end{frontmatter}

A serious limitation in the  field of polymer-based lasers
arises from the fact that all light emitting $\pi$-conjugated polymers
to date emit in the visible or UV. Telecommunications use infrared
radiation, so lasing at these wavelengths is desirable. Within conventional
theories of light emission from $\pi$-conjugated polymers,  
light emission in the IR from undoped $\pi$-conjugated polymers would be 
impossible. In this paper we point out the structural modifications that
can lead to emission in the IR.

Linear polyenes and trans-polyacetylene (t-PA) are
weakly emissive, because the lowest two-photon
state, the 2A$_g$, occurs below the optical 1B$_u$ state in these.
The optically pumped 1B$_u$ decays to the 2A$_g$ in ultrafast times, and
radiative transition from the 2A$_g$ to the ground state 1A$_g$ is forbidden.
Strong photoluminescence (PL) in systems like PPV and PPP implies 
excited state ordering E(2A$_g$) $>$ E(1B$_u$) (where E(...) is the energy of
the state), which is a consequence of enhanced effective bond alternation
within the effective linear chain model for
these systems \cite{Soos}. 
Since enhanced bond alternation necessarily
{\it increases} E(1B$_u$), it appears that strong PL is limited
to systems with optical gaps larger than that of t-PA.

Our goal is to demonstrate that materials obtained by
``site-substitution'' of t-PA, in which the hydrogen atoms of t-PA are replaced
with transverse conjugated groups will simultaneously have small optical
gaps {\it and} E(2A$_g$) $>$ E(1B$_u$). Initial work in this
area by us was limited to the specific system
poly-diphenylpolyacetylene (PDPA) \cite{Shukla,Ghosh}, in which the hydrogen
atoms of t-PA are replaced with phenyl groups.
Because of chain-bending
due to steric repulsion, PDPAs have short conjugation lengths \cite{Tada} and
emit in the visible. Further, although we determined from
our calculations that E(2A$_g$) $>$ E(1B$_u$) in PDPA,
because of the large number of atoms in the 
phenyl groups, our calculations were based on
uncontrolled approximations. Here we have chosen a hypothetical system for
which considerably improved many-body calculations can be done, and also, if
the system can at all be synthesized, steric repulsion would be minimal and
true long chain systems can be expected.
\begin{figure}
\begin{center}
\centerline{\includegraphics[width=5cm,draft=false]{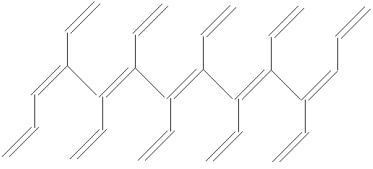}}
\caption{The hypothetical polyene investigated theoretically}
\end{center}
\end{figure}

Our calculations
below are for the substituted polyene shown in Fig.~1, with the hydrogen atoms
of a 10-carbon atom polyene replaced with ethylene. We
shall refer to this as the substituted polyene, and compare the excited state
ordering in this system with the ordinary 10-carbon atom polyene. Our 
calculations are within the dimerized Hubbard Hamiltonian
\begin{eqnarray}
&H&= -\sum_{j\sigma}t_j (c^\dagger_{j\sigma}
c_{j+1}+h.c) +
U \sum_j n_{j\uparrow} n_{j\downarrow}
\end{eqnarray}
where all terms have their usual meanings and $t_j$ = 2.2 eV and 2.6 eV for
single and double bonds, respectively. 
For both the polyene and the substituted
polyene we increase the Hubbard $U$ from zero (where E(2A$_g$) $>$ E(1B$_u$))
and determine the critical $U_c$ at which energy cross-over
E(2A$_g$) $<$ E(1B$_u$) occurs. For our speculation to be correct,
(i) E(1B$_u$, substituted) must be smaller than E(1B$_u$, unsubstituted),
and
(ii) $U_c$(substituted) must be greater than $U_c$(unsubstituted). The
latter would imply
that for a fixed $U$, E(2A$_g$, substituted) is higher 
than E(2A$_g$, unsubstituted).

Our calculations are based on the
exciton basis valence bond method \cite{Chandross}, 
within which the both the unsubstituted and substituted polyene
are considered as coupled two-level systems, where the two levels are the
HOMO and the LUMO of the unit cell ({\it i.e.} both calculations involve
10 MOs). This
approach is exact for the polyene, but 
approximate for the substituted polyene.
It is easy to prove that the 
$U_c$ calculated within the approximate method is a lower
limit for the true $U_c$(substituted). To prove this, we compare the exact
H\"uckel energes of the 10-carbon substituted polyene with those calculated
using the exciton basis.
The exact (approximate)
E(1B$_u$) = 1.02 (1.25) eV and E(2A$_g$) = 1.59 (1.65) eV. 
Now, since E(1B$_u$) increases with $U$, while E(2A$_g$) decreases (following
an initial weak increase),
it is seen from the above energies
that the $U$ at which the true E(1B$_u$) becomes larger than the true
E(2A$_g$) must be {\it higher} than that calculated within the reduced basis
set.
\begin{figure}[h]
\begin{center}
\centerline{\includegraphics[width=6cm,draft=false]{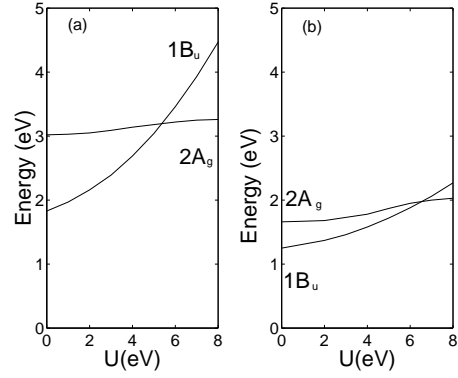}}
\caption{ E(1B$_u$) and E(2A$_g$) vs. $U$ for
the (a) unsubstituted and (b) the substituted polyene.} 
\end{center}
\end{figure}

In Figs. 2(a) and (b) we have plotted E(1B$_u$) and E(2A$_g$) against $U$ for
the unsubstituted and the substituted polyene. We note that (i)
E(1B$_u$, substituted) $<$ E(1B$_u$, unsubstituted) for any $U$, and (ii)  
$U_c$(substituted) is larger than $U_c$(unsubstituted) by $\geq$ 1 eV.
The reason why 
the 2A$_g$ (1B$_u$) is higher (lower) in energy in the substituted
system is as follows. In the strongly correlated electron model, the 1A$_g$
has all sites singly occupied with electrons, 
and the 1B$_u$ consists of a single doubly occupied site.
This double-occupancy in the
substituted polyene is delocalized over the entire transverse molecular unit,
and thus the {\it effective} Hubbard correlation, $U_{eff}$, is smaller. 
The lower $U_{eff}$ reduces E(1B$_u$) and
raises E(2A$_g$). Since the 1B$_u$ is predominantly a singly excited 
configuration, while the correlated 2A$_g$ consists of both single and double
excitations, one then expects that there is no qualitative difference 
between the unsubstituted and substituted 1B$_u$ wavefunctions at any $U$, 
but a strong
difference in the 2A$_g$ wavefunctions. This is exactly what is seen from our
calculations of the wavefunctions within the exciton basis.

Very similar results are obtained for N = 4,6 and 8 carbons on the backbone
polyene. 
Finite size scaling shows that considerably larger $U_c$ in the
substituted polyene is 
expected for N $\to \infty$. We conclude that
transverse site-substitution 
is one approach 
to obtain excited state orderings conducive to light emission in the IR.
Whether or not the present system can be
synthesized, we believe that the principle demonstrated here is 
general.

Work in Arizona was supported by NSF DMR-0101659 and NSF ECS-0108696.
Sandia is a multiprogram laboratory operated by Sandia Corporation, a
Lockheed Martin Company, for the United States Department of Energy under
Contract DE-AC04-94AL85000.


\begin{thebibliography}{99}
\bibitem{Soos} Z.G. Soos, S. Ramasesha, D.S. Galvao, Phys. Rev. Lett. 71
(1993) 1609.
\bibitem{Shukla} A. Shukla, S. Mazumdar, Phys. Rev. Lett. 83 (1999) 3944.
\bibitem{Ghosh} H. Ghosh, A. Shukla, S. Mazumdar, Phys. Rev. B 62 (2000)
12763.
\bibitem{Tada} K. Tada et. al., Proc. SPIE, 3145 (1997) 171.
\bibitem{Chandross} M. Chandross, Y. Shimoi, S. Mazumdar, Phys. Rev. B
59 (1999) 4822.
\end{thebibliography}
\end{document}